\documentclass[10pt,aps,prl,twocolumn,superscriptaddress]{revtex4-1}
\pdfoutput=1 
\usepackage{amsfonts}
\usepackage{mathrsfs}
\usepackage{amsmath}
\usepackage{amssymb}
\usepackage{fancyhdr}
\usepackage{graphicx}
\usepackage{xspace}
\usepackage{rotating}
\usepackage[normalem]{ulem}
\usepackage{braket}
\usepackage{verbatim}
\usepackage{xcolor}
\usepackage[utf8]{inputenc}
\usepackage[medium]{titlesec}
\usepackage{bm}
\usepackage[normalem]{ulem}
\usepackage{extarrows}
\usepackage{slashed}
\usepackage{isodateo}
\usepackage{graphicx}
\usepackage{xcolor}
\usepackage[bookmarksnumbered=true,bookmarksopen=true]{hyperref}
\usepackage[hmargin=.7in,vmargin=1.1in]{geometry}
\usepackage{indentfirst}
\usepackage{cancel}

\graphicspath{{./}{./Figs/}{./figs/}{../Figs/}{./fig/}}

\usepackage{amsfonts}
\usepackage{mathtools}
\usepackage{pifont}
\usepackage{mathrsfs}
\usepackage{amsmath}
\usepackage{amssymb}
\usepackage{framed}

\newcommand{\nc}{\newcommand}

\newcommand{\bit}{\begin{itemize}}  
\newcommand{\eit}{\end{itemize}}

\newcommand{\benum}{\begin{enumerate}}
\newcommand{\eenum}{\end{enumerate}}
\newcommand{\bi}{\begin{itemize}}
\newcommand{\ei}{\end{itemize}}
\newcommand{\Mp}{M_{\rm pl}}
\newcommand{\mpl}{M_{\mathrm{Pl}}}

\newcommand{\beq}{\begin{equation}}
\newcommand{\eeq}{\end{equation}}

\newcommand{\bea}{\begin{eqnarray}}
\newcommand{\eea}{\end{eqnarray}}

\newcommand{\Rmnum}[1]{\expandafter\@slowromancap\romannumeral #1@}

\def\bga{\begin{aligned}}
\def\eda{\end{aligned}}
\def\bgp{\begin{pmatrix}}
\def\edp{\end{pmatrix}}
\def\bgs{\begin{subequations}}
\def\eds{\end{subequations}}

\def\mf{\mathfrak}

\def\to{\rightarrow}

\def\Mp{M_{\text{Pl}}}
\def\mphi{m_{\Phi}}
\def\lphi{\lambda_{\Phi}}

\def\mf{m_{_\Phi}}
\def\lf{\lambda_{_\Phi}}
\def\phiin{\phi_{\mathrm{in}}}
\def\phiR{\phi_{_\mathrm{R}}}
\def\phiI{\phi_{_\mathrm{I}}}
\def\phiRIn{\phi_{_\mathrm{R,in}}}
\def\phiIIn{\phi_{_\mathrm{I,in}}}
\def\absPhi{\left|\Phi\right|}
\def\mR{m_{_\mathrm{R}}}
\def\mI{m_{_\mathrm{I}}}
\def\Gphi{\Gamma_{\Phi}}
\def\gphi{\gamma_{\Phi}}

\renewcommand{\rm}{\mathrm}

\def\phiR{\phi_{_\mathrm{R}}}
\def\phiI{\phi_{_\mathrm{I}}}
\def\phiRS{\phi_{_\mathrm{R,\ast}}}
\def\phiIS{\phi_{_\mathrm{I,\ast}}}
\def\phiRIn{\phi_{_\mathrm{R,in}}}
\def\phiIIn{\phi_{_\mathrm{I,in}}}
\def\phiIn{\phi_{_\mathrm{in}}}
\def\phiS{\phi_{_\ast}}

\def\mf{m_{_\Phi}}
\def\lf{\lambda_{_\Phi}}
\def\mR{m_{_\mathrm{R}}}
\def\mI{m_{_\mathrm{I}}}
\def\absPhi{\left|\Phi\right|}
\def\nc{n_{_C}}
\def\Gphi{\Gamma_{\Phi}}
\def\gphi{\gamma_{\Phi}}

\def\nct{\tilde{n}_{_C}}
\def\Mp{M_{\mathrm{P}}}

\usepackage{eso-pic}
\newcommand{\reportnum}[2]{
  \AddToShipoutPictureBG*{%
    \AtPageUpperLeft{%
      \hspace{0.75\paperwidth}%
      \raisebox{#1\baselineskip}{%
        \makebox[0pt][l]{\textnormal{#2}}
  }}}%
}

\begin{document} 

\reportnum{-3}{KEK-TH-2672}
\reportnum{-4}{KEK-Cosmo-0366}

\title{Probing baryogenesis with gravitational waves}

\author{Yanou Cui}
\email[]{yanou.cui@ucr.edu}
\affiliation{Department of Physics and Astronomy, University of California, Riverside, CA 92521, USA}

\author{Anish Ghoshal}
\email{anish.ghoshal@fuw.edu.pl}
\affiliation{Institute of Theoretical Physics, Faculty of Physics, University of Warsaw,ul.  Pasteura 5, 02-093 Warsaw, Poland}

\author{Pankaj Saha}
\email{pankaj@post.kek.jp}
\affiliation{Institute of Particle and Nuclear Studies~(IPNS), High Energy Accelerator
Research Organization (KEK), Oho 1-1, Tsukuba 305-0801, Japan}

\author{Evangelos I. Sfakianakis}
\email{evangelos.sfakianakis@austin.utexas.edu}
\affiliation{
Texas Center for Cosmology and Astroparticle Physics,
Weinberg Institute for Theoretical Physics, Department of Physics,
The University of Texas at Austin, Austin, TX 78712, USA}
\affiliation{Department of Physics, Harvard University, Cambridge, MA, 02131, USA}
\affiliation{Department of Physics, Case Western Reserve University, Cleveland, OH 44106, USA}

\date{\today}

\begin{abstract}
Affleck-Dine (AD) baryogenesis is compelling yet challenging to probe because of the high-energy physics involved.
We demonstrate that this mechanism can be generically realized with low-energy new physics without supersymmetry while producing detectable gravitational waves (GWs) sourced by parametric resonance of a light scalar field.
In viable models, the scalar has a mass of $\mathcal{O}(0.1-10)$ GeV,  yielding GWs with peak frequencies of $\mathcal{O}(10-100)$ Hz.
This study further reveals a new complementarity between upcoming LIGO-frequency GW detectors and laboratory searches across frontiers of particle physics.

\end{abstract}
\maketitle

\noindent\textbf{Introduction.}
The observed excess of matter over antimatter in the universe, in terms of the baryon-to-photon ratio $\eta = \frac{n_{b}-n_{\bar{b}}}{n_{\gamma}} \approx 6 \times 10^{-10}$ \cite{Aghanim:2018eyx}, remains a long-standing puzzle in particle physics and cosmology. Affleck-Dine (AD) baryogenesis~\cite{Affleck:1984fy} is a compelling solution to the puzzle, where the asymmetry emerges from the oscillation and decay of a baryon number-carrying complex scalar field. Originally formulated with supersymmetry (SUSY)~\cite{Affleck:1984fy,Dine:1995kz,Dine:2003ax,Enqvist:2003gh}, it involves flat direction moduli fields composed of superpartners of standard model (SM) particles. These light moduli fields can acquire large displacements during inflation, break C and $CP$ symmetries and subsequently roll down and oscillate to produce baryon asymmetry, while the inflaton still dominates the Universe~\cite{Allahverdi:2012ju} (alternatively, the AD field itself can be the inflaton~\cite{Hertzberg:2013mba,Cline:2019fxx}).
\par
Despite its appeal, experimentally testing the AD mechanism is challenging because of the high-scale physics involved in the inflationary era. 
Fortunately, the stochastic gravitational wave background (SGWB) generated by the oscillating AD field offers a promising new probe by utilizing the naturally high-energy environment in the early cosmos: oscillating scalar fields can trigger \textit{parametric resonance}, i.e., rapid nonperturbative particle production, leading to large inhomogeneities and fragmentation
that source GWs (e.g.,~\cite{Kitajima:2018zco, Figueroa:2017vfa, Lozanov:2019ylm, Hiramatsu:2020obh, Easther:2006vd, Friedberg:1976me,Coleman:1985ki,Kusenko:1997ad, Kusenko:1997si}).
\par
In conventional SUSY AD baryogenesis, parametric resonance does not generate detectable GWs~\cite{Kusenko:2008zm,Kusenko:2009cv,Zhou:2015yfa}~
\footnote{
In SUSY AD models, producing detectable GWs can typically lead to $\eta_B\gg 10^{-10}$~\cite{Kusenko:2008zm, Kusenko:2009cv}. Therefore,~\cite{Kusenko:2008zm, Kusenko:2009cv} considered GWs from AD flat-direction fields with $B\!-\!L = 0$,  not relevant for baryogenesis.
Furthermore, the potential term $|\Phi|^2\log (|\Phi| / M_{\rm {Pl}})$  considered in \cite{Kusenko:2008zm, Kusenko:2009cv, Zhou:2015yfa} exhibits a pathology at $\Phi=0$, which is absent in our polynomial potential; its impact on GW emission warrants re-examination".
} 
Meanwhile, SUSY AD models can lead to Q-balls, which may result in early matter domination and induce GWs at second order by enhancing preexisting curvature perturbations~\cite{Kusenko:2008zm, Kusenko:2009cv}. Q-ball-induced GWs strongly depend on the speed of the transition from matter to radiation domination~\cite{White:2021hwi}.

The purpose of this work is twofold: first, we present non-SUSY realizations of the AD mechanism that involve new physics below the electroweak scale and can operate in cosmic epochs long after inflation; second, we demonstrate that parametric resonance sourced by AD condensates in such models naturally produces SGWB within reach of upcoming GW experiments.

This study uncovers a generic class of well-motivated sources of SGWB and adds a timely complement to the rapidly growing field of probing particle physics with GWs, building on the ground-breaking discoveries of GWs by LIGO-Virgo-KAGRA collaboration~\cite{LIGOScientific:2016aoc,LIGOScientific:2016sjg,VIRGO:2014yos} and the recent strong evidence for SGWBs revealed by pulsar timing arrays~\cite{Carilli:2004nx,Weltman:2018zrl,EPTA:2015qep,EPTA:2015gke,NANOGrav:2023gor,NANOGrav:2023hvm, EPTA:2023sfo,EPTA:2023fyk}.
Based on simple benchmark models, we show that successful baryogenesis requires new physics in the AD sector with a characteristic scale $m_{\Phi}={\cal O}(0.1-10)$ GeV, and the resulting  GWs have a peak frequency in the range of ${\cal{O}}(1-100)$ Hz, within reach of upcoming detectors, including the Cosmic Explorer
(CE) \cite{LIGOScientific:2016wof} and Einstein Telescope (ET) \cite{Hild:2008ng}, as well as DECIGO \cite{Kawamura:2020pcg} and Big Bang Observatory (BBO)~\cite{Harry:2006fi} (for the low-frequency tail).
The study further uncovers a new potential complementarity between GW detection and laboratory
experiments at energy, intensity, and neutrino
frontiers, such as the LHC, DUNE, SHiP, and FASER \cite{ATLAS:2020syg, ATLAS:2018umm, Bolton:2019pcu,FASER:2018eoc,Ballett:2019bgd,T2K:2019jwa,Drewes:2018gkc,Barouki:2022bkt}, in search for new physics, depending on the details of the asymmetry transfer interactions between the AD field and the SM.

\medskip

\noindent\textbf{Benchmark Models.}
\noindent\textit{I. AD scalars: background dynamics and asymmetry generation}\\
We consider a complex scalar field $\Phi$ (SM gauge singlet) with canonical kinetic terms, Einstein gravity, and a simple renormalizable polynomial potential:
\begin{equation}
    V(\Phi) = \lphi\absPhi^4 + \mphi^2\absPhi^2 - A(\Phi^n + \Phi^{\ast n}) \label{eq:V},
\end{equation}
where $n \le 4$ is an integer. 
We choose $n=2$ for concreteness.  
Although other choices may lead to similar results, $n=2$ stands out as a simple and natural option ~\footnote{For example, with the alternative choice of $n = 4$, the generated asymmetry quickly freezes in at a large value,
requiring a very small value of A to reproduce the observable baryon asymmetry\cite{Rubakov:2017xzr, Hertzberg:2013mba}. In contrast, with
a quadratic symmetry-breaking term (n = 2), the oscillations of the $\Phi$-condensate around the origin continue
to generate and cancel asymmetry over time, partially suppressing the initially large asymmetry~}.
The first two terms manifest a $U(1)$ symmetry, which is broken by the third term as necessary to generate the $\Phi$-$\Phi^{\dagger}$ asymmetry, precursor to the baryon asymmetry. 
We consider the general possibility that $\Phi$, as a light spectator during inflation, starts with a large (even Planckian) field displacement \cite{Starobinsky:1994bd, Cui:2023fbg, Hardwick:2017fjo, Markkanen:2018gcw, Markkanen:2019kpv}, for instance, due to de Sitter fluctuations during inflation. We have verified that the isocurvature constraints from CMB data, which are relevant for any light spectator field, can be satisfied.
We denote the typical value at the end of inflation as $\Phi_{\rm {in}}$, which is the initial value for our calculations~
\footnote{
Other mechanisms can drive the $\Phi$ field to large values during inflation, like a non-minimal coupling to gravity or a direct coupling to the inflaton. 
Our results for GWs and $n_b-n_{\bar b}$ do not depend on the origin of $\Phi_{\rm{in}}$.
}.
As long as the (radial) field is light, $H\gg H_{\rm {osc}}\simeq \sqrt{V_{,|\Phi|}/|\Phi|}$, where $H_{\rm{osc}}$ is the effective mass of $|\Phi|$~\cite{Cui:2023fbg}, it remains frozen
\footnote{Since we consider small $U(1)$ symmetry-breaking terms, the angular field will always be light when the radial field is light.}.
The initial phase of $\Phi$, $\theta=Arg(\Phi)$, is also set as $\theta_{\rm {in}}$ at the end of inflation in our observable patch of the Universe. As the Universe expands, the Hubble rate drops, and at $H\simeq H_{\rm {osc}}$ the field starts rolling.
For our parameters, the rolling starts during the radiation-dominated (RD) era, well after the end of inflation.

The real and imaginary parts of $\Phi\equiv \frac{1}{\sqrt{2}}(\phi_R + i \phi_I)$ obey the following equations of motion:
\begin{align}
\label{eq:phiR}
\ddot{\phi}_{_\mathrm{R}} &+ 3H\dot{\phi}_{_\mathrm{R}} + \mR^2\phiR + \lphi(\phiR^2 + \phiI^2)\phiR = 0,\\
\label{eq:phiI}
\ddot{\phi}_{_\mathrm{I}} &+ 3H\dot{\phi}_{_\mathrm{I}}~ + \mI^2\phiI~ + \lphi(\phiR^2 + \phiI^2)\phiI~ = 0,
\end{align}
where $m^2_{R,I} = m_\Phi^2 \mp 2A$.
The Hubble expansion rate in an RD Universe is $H\propto t^{1/2}$. An overdot denotes derivative with respect to cosmic time $t$.
If $A\neq 0$, the real and imaginary components have different masses, so their trajectories in the $\phi_R$-$\phi_I$ plane resemble shrinking ellipses (not straight lines), even for zero initial velocities.

The background field evolution (assuming $A\ll \mphi^2$) is initially dominated by the $\absPhi^4$ term, where the amplitude scales as $\phi\propto 1/a$. When the amplitude drops below $|\Phi_{\ast}| = \mphi/\sqrt{\lphi}$ at $t=t_*$, the $\absPhi^2$ term dominates the oscillation and the
fields evolve approximately as
$\phi_{R,I} \sim a^{-3/2}\cos(m_{R,I} t +c_{R,I})$, with $c_{R,I}$ being a constant phase
(see Supplemental Material, Section C, for details).
The $\Phi$-$\Phi^\dag$ asymmetry, which will be subsequently transferred to the baryon (or lepton) number, is
$
    n_\Phi(t) \equiv i\left(\Phi^{\dagger}\dot{\Phi} - \dot{\Phi}^{\dagger}\Phi\right)
         = \dot{\phi}_{\rm R}\phiI - \phiR\dot{\phi}_{\rm I}
$. In the absence of significant washout,
$n_\Phi$ directly corresponds to $n_B$ or $n_L$ (see the Discussion later). Using the above approximations for $\phi_{R,I}$ and including the terms $\Gamma_\Phi \dot \phi_{R,I}$ to
model the decay of $\Phi$ to SM states (the specific form of $\Gphi$ depends on the decay channels, which we elaborate on later), the comoving baryon asymmetry generated during the time $t>t_*$ is~\cite{Rubakov:2017xzr,Lloyd-Stubbs:2020sed,Lloyd-Stubbs:2022wmh,Mohapatra:2021aig}:
\begin{align}
\nonumber
    &\left(\frac{a(t)}{a_{\mathrm{in}}}\right)^3n_{B}(t) \simeq 4A\phiRIn\phiIIn\left(\frac{\phiin}{\phi_*}\right)\times\\
    &\int_{t_*}^{t}dt'\cos{(\mR(t'-t_\ast))}\sin{(\mI(t'-t_\ast))}e^{-\Gphi(t'-t_*)},
    \label{eq:com_nb}
\end{align}
which can be further simplified for $A\ll m_\phi^2$  (the detailed derivation is presented as Supplemental Material) and scales as
\begin{align}
\label{eq:asymmetryintegral}
\frac{a^3(t)}{a^3_{\mathrm{in}}}n_{B}(t) \propto 
\int_{t_*}^t dt'  \sin\left(
\frac{2A}{m_\Phi} (t-t_*)
\right) e^{-\Gamma_\Phi(t-t_*)}
\,.
\end{align}
Evaluating the above integral for $t\to\infty$ and factoring in the expansion during the RD Universe, we express $n_B/s$ as
\begin{align}
    \frac{n_B}{s} =
    \begin{cases}
    \left(\frac{4\alpha^3}{\lphi k_{T_d}^6}\right)^{1/4} \epsilon_\Phi\frac{\mphi M_{\rm {Pl}}}{T_d^2}\sin(2\theta);&\gphi \gg 2\epsilon_\Phi\\
  \left(\frac{\alpha^3k_{T_d}^2}{64\lphi}\right)^{1/4}\frac{1}{\epsilon_\Phi}\frac{T_d^2}{\mphi M_{\rm{Pl}}}\sin(2\theta); &\gphi \ll 2\epsilon_\Phi
    \end{cases}
    \label{eq:nbsmainDmnless}
\end{align}
where $M_{\rm{Pl}} \simeq 2.4\times 10^{18}$ GeV is the reduced Planck mass, and
we define the dimensionless quantities: $\epsilon_{\Phi}=A/\mphi^2$, $\gamma_{\phi}=\Gphi/\mphi$ and $k_{T_d} = \sqrt{\pi^2g(T_d)/90}$.
$T_d$ is the temperature of the Universe characterizing $\Phi$ decay, as defined by $H(T_d) =\Gphi$, $\alpha$ is the initial fraction of energy density in the spectator sector $\Phi$ relative to the total energy density, and $\theta = \theta_{\rm {in}}$ is defined earlier. 
The exact form of $n_B/s$, from which Eq.~\eqref{eq:nbsmainDmnless} is derived, is given as Supplemental Material.

The change in the scaling of the final asymmetry with the normalized symmetry–breaking parameter $\epsilon_\Phi \equiv A/m_\Phi^2$ shown by the two branches of Eq.~\eqref{eq:nbsmainDmnless} can be understood simply in terms of the competition between two timescales: the oscillation period of the asymmetry and the lifetime of the scalar. 
The symmetry–breaking parameter $\epsilon_\Phi$  induces a mass splitting between the particle and antiparticle components of $\Phi$, leading to an oscillation of the charge asymmetry with frequency $\omega_\Phi \sim \Delta m_\Phi \sim \epsilon_\Phi m_\Phi$ and period $T_{\mathrm{asy}} \sim 1/(\epsilon_\Phi m_\Phi)$. 
At the same time, the scalar decays with rate $\Gamma_\Phi$, corresponding to a lifetime $\tau_\Phi \sim 1/\Gamma_\Phi$. 
The net asymmetry transferred to the thermal bath is proportional to the time integral of an oscillatory source with an exponential envelope, schematically   $n_B\propto\int_0^\infty dt e^{-\Gamma_\Phi t} \sin(\omega_\Phi t)$, so the final result depends on whether the decay occurs before or after many oscillations. 
In the regime $\Gamma_\Phi \gg \omega_\Phi$ (i.e. $\tau_\Phi \ll T_{\mathrm{asy}}$), the integral is dominated by very early times, when the sine function has not yet completed a full oscillation and may be approximated as $\sin(\omega_\Phi t) \simeq \omega_\Phi t$. 
The generated asymmetry is then essentially fixed before sign–flipping oscillations set in, and scales linearly with the oscillation frequency (equivalently with the asymmetry parameter $\epsilon_\Phi$) as
\beq
n_B \propto \int_0^{\sim 1/\Gamma_{\Phi}}dte^{-\Gamma_{\Phi}t}\omega_\Phi t\sim \frac{\omega_\Phi}{\Gamma_\Phi}  \propto \epsilon_\Phi.
\eeq
By contrast, in the regime $\Gamma_\Phi \ll \omega_\Phi$ (i.e. $\tau_\Phi \gg T_{\mathrm{asy}}$), the source oscillates many times before the field decays, spending comparable time with positive and negative signs. 
As a result, contributions from successive oscillations partially cancel, and the net asymmetry is a residue arising from the slow decay of the envelope. To see the overall scale of the asymmetry for this limit, we note that the contribution from a single half-period of the oscillation is of order $\int dt\sin(\epsilon_\Phi m_\Phi t) \sim 1/(\epsilon_\Phi m_\Phi)$, which sets the characteristic scale of the integral in Eq.~\eqref{eq:asymmetryintegral}. 
Including all subsequent oscillations leads to an alternating geometric series, which can be summed and consequently multiplies the result of the first oscillation by $1/2$ for the case with $\gamma_\phi/2\epsilon_\Phi\ll 1$. 
The final asymmetry thus scales as
\begin{align}
\nonumber
n_B \propto & \int_0^{\infty}dt \sin(\epsilon_\Phi m_\Phi t)\\
\simeq&
\frac12\times\int_0^{\pi/(2\epsilon_\Phi m_\Phi))}dt \sin(\epsilon_\Phi m_\Phi t)
\sim \frac{1}{2}\frac{1}{\epsilon_\Phi m_\Phi}.
\end{align}
In summary, a quadratic symmetry-breaking term induces a constant mass splitting between the real and imaginary components of $\Phi$, leading to a single well–defined oscillation frequency $\omega_\Phi \propto \epsilon_\Phi m_\Phi$. 
This simple harmonic structure makes the final asymmetry an explicit function of $\omega_\Phi$, and hence of $\epsilon_\Phi$, producing the characteristic two-regime behavior\footnote{Note that the $\epsilon_\Phi\to 0$ limit is only reached through the upper branch of Eq.~\eqref{eq:nbsmainDmnless}, leading to $n_B/s\to 0$ as expected.} $n_B \propto \epsilon_\Phi$ for $\Gamma_\Phi \gg \omega_\Phi$ and $n_B \propto 1/\epsilon_\Phi$ for $\Gamma_\Phi \ll \omega_\Phi$, shown in Eq.~\eqref{eq:nbsmainDmnless}.
In the case where the asymmetry in the $\Phi$ field is first transferred to an SM lepton asymmetry $n_L$, which is later converted to a baryon asymmetry $n_B$ via the sphaleron process, the above result gets multiplied by a factor of $28/79$ since~\cite{Harvey:1990qw,Asaka:2002zu, Garcia-Bellido:1999xos}
\(n_B/s = -(28/79)\times n_L/s\,.\)

\indent Equation~\eqref{eq:nbsmainDmnless} is derived assuming that the asymmetry is generated when the potential is dominated by the $|\Phi|^2$ term, thus requiring that the background dynamics prior to $t_*$ 
is not affected by the $A$ term. The corresponding condition on $A$ is derived as follows: the only term in the potential that involves the angular degree of freedom is the symmetry-breaking term $2 A|\Phi(t)|^2 \cos 2 \theta(t)\subset V(|\Phi|,\theta) $, indicating an upper bound on the effective mass along the angular direction: $|m^2_\theta|\leq 2 A$. The angular degree of freedom is frozen at $\theta_{\rm {in}}$ for $t<t_*$ if $|m^2_\theta|<H^2(t_*)$~\cite{Lloyd-Stubbs:2022wmh}, which translates into $\epsilon_\Phi \equiv A/\mphi^2 < ({\mu^2}/{16\alpha})\left(\phi_{\rm {in}}^2/M_{\rm{Pl}}^2\right)$, where $\mu^2 \equiv \mphi^2/(\lphi\phiin^2)$, and $\alpha$ is defined below Eq.~\ref{eq:nbsmainDmnless}. As we see in the next section, ${\phiin} \simeq\mpl$ and $\mu^2\lesssim 10^{-5}$ are required for detectable GW signals.
With larger $A$, motion along the angular direction is non-negligible, and $\theta(t)$ would have already evolved toward its minimum before $t=t_*$, thus reducing $\sin\theta$ and suppressing the asymmetry [see Eq.~\eqref{eq:nbsmainDmnless}].
For our benchmark examples, the chosen $A$ values satisfy this self-consistency condition.

We now present some numerical examples of the asymmetry. For $\epsilon_\Phi=10^{-20}$ the observed $n_B/s\sim 10^{-10}$ can be realized with $T_d\sim 2.5\times 10^7$ GeV when $\mphi=10\,\mathrm{GeV}$. As we will discuss, transferring the asymmetry to the baryon sector may involve a significant (model-dependent) washout, requiring a larger initial $\Phi$ asymmetry to compensate. Therefore, we also consider parameter points that lead to $n_\Phi/s\gg 10^{-10}$, for example, the same $\epsilon_\Phi$ with $T_d\sim 1$ GeV gives $n_\Phi/s \sim 4\times 10^{-5}$.
A more detailed parameter scan is provided in Section C of Supplemental Material, demonstrating the potential to produce a large initial asymmetry that compensates for possible washout effects.

\medskip

\noindent \textit{II. Transfer to the SM baryon asymmetry~}.\\
The asymmetry in the $\Phi$ field generated during its oscillation needs to eventually be transferred to the SM baryon asymmetry via its subsequent decay to baryons or leptons (further assisted by electroweak sphalerons in the latter case). This can generally be realized by a B- or L-violating coupling involving $\Phi$ in terms of effective higher-dimensional operators, such as $\Phi QQQL/\Lambda^3$~\cite{Allahverdi:2012ju}, $\Phi UDDUDD/\Lambda^6$, and $\Phi LLELLE/\Lambda^6$. Whether the related constraints, for instance, from $n-\bar{n}$ oscillation, can be satisfied depends on the specifics of the UV completion, including flavor structure in the couplings. As a proof of principle, we outline examples of renormalizable, UV-complete models that realize the $\Phi$ to baryon asymmetry transfer in the following, and summarize the conditions on viable model parameters. Further details are given in Section D of Supplemental Material.
The core results of this work are independent of the details of the asymmetry transfer mechanism.\\
\textit{(A) $\Phi$ decay to leptons (leptogenesis).}\\
A simple renormalizable model involves:
\beq
\mathcal{L}\supset y_{N,~ i}\Phi\bar{N_i}^C{N_i}+g_{\nu,~ij}\bar{N}_{i}HL_j+ {\rm{h.c.}},
\eeq
where $N_i$'s are right-handed Majorana neutrinos (with $i=1,2,3$ as a typical choice), $H$ is SM Higgs, and $L_j$ are SM lepton doublets. This model potentially relates to neutrino mass generation, which, nevertheless, is not imposed as a requirement here. We consider the simple scenario where $\Phi$ decays to $NN$, then $N$ transfers the asymmetry to SM leptons via freeze-out or freeze-in processes enabled by the $NHL$ coupling, as considered in the literature of leptogenesis \cite{Akhmedov:1998qx, Klaric:2020phc, Klaric:2021cpi, Cui:2011qe, Hambye:2016sby, Flood:2021qhq} \footnote{Unlike in leptogenesis, here we do not require additional CP violation or out-of-equilibrium condition in neutrino sector to generate the asymmetry, as the asymmetry is already generated in $\Phi$.}.

\noindent \textit{(B) $\Phi$ decay to baryons (direct baryogenesis)}\\
A simple renormalizable model, in this case, includes
\beq
\mathcal{L}\supset y_\chi\Phi\chi\chi+g_i\chi\bar{u}_i\varphi+\lambda_{ij}\varphi d_i d_j +{\rm{H.c.}},
\eeq
where $\chi$ is a singlet Majorana fermion, $\varphi$ is an up-type diquark scalar, and the flavor indices $i,~j=1,2,3$ in down-type quark combination $d_i d_j$ should be antisymmetric. In the viable parameter region, the decay proceeds with $\Phi\rightarrow \chi\chi$, followed by $\chi\rightarrow udd$.

Both models are subject to cosmological conditions that impose lower limits on $m_\Phi$, $m_\chi$, or $m_N$. For $\Phi$ to decay to baryons (before BBN), it is subject to kinematic conditions: $m_\Phi\gtrsim 2$ GeV, and $m_\chi\gtrsim1$ GeV. For $\Phi$ asymmetry to be transferred to SM leptons via $N$ before the EW phase transition requires $m_\Phi>2 m_N\gtrsim O(0.1)$ GeV \cite{Klaric:2020phc}. In addition, for both models, the potential asymmetry washout effects should be either suppressed or compensated by a larger initial $\Phi$ asymmetry produced during its oscillation (which has been shown to be viable in the last section). Furthermore, relevant laboratory experiments at the energy and intensity frontiers, searching for B-violation, color-charged new particles, and sterile neutrinos, among others, may provide additional constraints on model parameters. In Section D of Supplemental Material, we elaborate on relevant constraints and demonstrate examples of viable parameter regions. 

\medskip

\noindent\textbf{Parametric Resonance and GW Signals.}
We have seen that the generation of the baryon asymmetry is determined by the classical motion of $\Phi$ as an averaged background field.
In addition to this classical motion of $\Phi$, the corresponding fluctuations around it can lead to parametric resonance and source GWs.
We decompose $\phi_{R,I} = \bar\phi_{R,I}(t) + \delta\phi_{R,I}(x,t)$, where $\bar\phi_{R,I}$ describe the classical background motion and $\delta\phi_{R,I}(x,t)$ correspond to the fluctuations, which have a quantum origin.
The evolution of background fields $\bar\phi_{R,I}$ follows Eqs.~\eqref{eq:phiR}~and~\eqref{eq:phiI}, whereas we expand the fluctuations in Fourier modes as
\beq
\delta\phi_{R}(x,t) = \int {d^3k\over (2\pi)^3} \left[
a_{R,k} e^{i\vec k \cdot \vec x} \delta\phi_{R,k}(t) + H.c.
\right ]
 \eeq
and similarly for $\delta\phi_I(x,t)$.

As the background fields oscillate around the minimum of the potential, the fluctuations in momentum space evolve according to
\begin{eqnarray}
\nonumber
&&\ddot{\delta \phi}_{R,I} + 3 H \dot {\delta \phi}_{R,I} + m_{R,I}^2{\delta \phi_{R,I}} + {k^2\over a^2}  {\delta \phi_{R,I}} \\
&&\quad + \lambda_\phi (
3 \bar\phi_{R,I}^2  + \bar \phi_{I,R}^2)\delta\phi_{R,I}
+ 2 \bar\phi_I \bar\phi_R \delta \phi_{I,R}=0.
\end{eqnarray}
We thus see that, in general, the equations for $\delta\phi_{R,I}$ are coupled. We first consider the representative limit of $A=0$, where $m_R=m_I=m_\Phi$ and the background field trajectory is simply a straight line through the origin (for vanishing initial velocities, as is the case here).

In this case, we can rotate our field basis and define fields in the radial direction (along the background motion) $\phi_\parallel$ and perpendicular to it (orthogonal direction) as $\phi_\perp$. By construction, $\bar \phi_\perp=0$, and the fluctuation equations for $\phi_\parallel$ and $\phi_\perp$ decouple, namely,
\begin{eqnarray}
\nonumber
\ddot {\delta\phi}_{\parallel,\perp} + 3 H \dot{\delta\phi}_{\parallel,\perp}
+  {k^2\over a^2}  {\delta\phi}_{\parallel,\perp}
+ m_\Phi^2 {\delta\phi}_{\parallel,\perp}
\\+ q_{\parallel,\perp} \lambda_\Phi \bar \phi_{\parallel,\perp}^2  \delta\phi_{\parallel,\perp}=0,
\end{eqnarray}
where $q_{\parallel,\perp}=\{3,1\}$.
Quasiperiodic background-field oscillations can exponentially amplify fluctuations within certain wave-number ranges. This rapid growth of mode occupation numbers allows these modes to be treated classically \cite{Polarski:1992dq, Kofman:1997yn}. This phenomenon, originally studied for exponential particle production (``preheating") after inflation~\cite{Greene:1997fu}, can also occur for any oscillating scalar (see, for instance, \cite{Cui:2023fbg}).

To quantify parametric resonance, we define a dimensionless field amplitude $\tilde{\phi} = \phi/\phiin$ and a corresponding conformal scaling for time as $\mathrm{d}\eta = \sqrt{\lphi}\phiin\mathrm{d}t/a$ with $\phiin$ being the initial amplitude of the scalar field $\phi_{\rm {in}} = \left |\Phi_{\rm {in}}\right |$.
The dimensionless fluctuation equation becomes ($\delta\phi'\equiv d\delta\phi/d\eta$):
\begin{equation}
 {\delta \phi}_{\parallel,\perp}'' + 2 \tilde H  {\delta \phi}_{\parallel,\perp}'
+ \left ( {\tilde k^2}
+ a^2 \mu^2  + q_{\parallel,\perp} a^2 \tilde \phi^2  \right ) \delta\phi_{\parallel,\perp}=0,
\end{equation}
where $\tilde k^2 = k^2 / \lambda \phi_{\rm{in}}^2$ and
$\tilde \phi  = \bar\phi / \phi_{\rm {in}}$.

There are two essential resonance parameters:
$\mu^2 = \frac{\mphi^2}{\lphi\phiin^2}$
and $q_{\parallel , \perp}$. As shown in \cite{Greene:1997fu}, a large $\mu^2$ can suppress the resonance and determine if the system displays
 \emph{stable resonance}, \emph{stochastic resonance}, or \emph{no decay}.
We choose to focus on $\mu \leq 10^{-5}$, such that the system is always in a stable resonance regime. For such small $\mu^2$, the system is similar to the well-studied case of a quartic field during preheating, whereas $q_{\parallel, \perp}$ is fixed in our case due to the rotational symmetry of the potential for $A=0$.
Section E of Supplemental Material discusses in detail the effect of $q_{\parallel,\perp}$ on the growth of $\delta\phi_{\parallel,\perp}$ and shows that for sufficiently small values of the asymmetry parameter $A \lesssim 0.005 \mphi^2$, the parametric resonance dynamics is indistinguishable from the case $A=0$, and thus the GW amplitude remains insensitive to the value of $A$. Fortunately, the parameter range of $A$ that produces sufficient baryon asymmetry, as considered in our benchmarks, is
well within this regime.
On a technical note, we use the parametrization of real and imaginary components of $\Phi$ ($\phi_R$, $\phi_I$) when computing the baryon asymmetry, whereas we are using the radial direction and its orthogonal ($\phi_\parallel$, $\phi_\perp$) when computing parametric resonance for $A\ll \mphi^2$. This is due to the fact that the baryon asymmetry scales with $A$ and thus the difference between the $\phi_I$ and $\phi_R$ directions is important, whereas parametric resonance and GW production are blind to $A\ll m_\Phi^2$, making radial coordinates more suitable. Interesting deviations from the $A=0$ limit are discussed in Section E of Supplemental Material. For $A\gtrsim 0.05m_\Phi^2$ the GW spectrum starts being sensitive to the exact value of $A$ as well as the initial angle $\theta_{\rm{in}}$. This provides intriguing possibilities for distinguishing between these parameters by dedicated analysis of the GW spectral shape.

The growth of fluctuations and their backreaction breaks the homogeneous oscillation of $\Phi$, leading to inhomogeneities with an effective anisotropic stress tensor $\Pi_{ij}^{TT} = \nabla_i\phi^a\nabla_j\phi^a$ (where the repeated index $`a$' indicates the number of the scalar components and a summation is implied).
The dynamics of the generated GWs are governed by the linearized equation for the transverse-traceless tensor perturbation sourced by the above anisotropic stress
\begin{equation}
    \ddot{h}_{ij} + 3H\dot{h}_{ij} - \frac{\nabla^2h_{ij}}{a^2} = \frac{2}{a^2\Mp^2}\Pi_{ij}^{TT} \, .
    \label{eq:Eqhij}
\end{equation}
The quantity of interest is the energy density power spectrum of the GWs normalized by the critical energy density of the Universe today.
For GWs produced in a radiation-dominated era, the spectrum of the energy density of GWs (per logarithmic momentum interval) observable today is~\cite{Easther:2006gt,Easther:2006vd,Easther:2007vj,Dufaux:2007pt,Cui:2023fbg, Garcia-Bellido:2007nns, Garcia-Bellido:2007fiu}
\begin{align}
\nonumber
\Omega_{\rm GW,0}h^2 &= \frac{h^2}{\rho_{\mathrm{crit}}}\frac{d\rho_{\rm GW}}{d\ln k}\Bigg|_{t=t_0} = \frac{h^2}{\rho_{\mathrm{crit}}}\frac{d\rho_{\rm GW}}{d\ln k}\Bigg|_{t=t_\mathrm{e}}\frac{a_\mathrm{e}^4\rho_{\mathrm{e}}}{a_0^4\rho_{\mathrm{crit},0}}\\
&= \Omega_{\mathrm{rad}, 0}h^2\Omega_{\mathrm{GW},\mathrm{e}}\left(\frac{g_{\ast}}{g_0}\right)^{-1/3},
\end{align}
where the critical density of the Universe is $\rho_{\mathrm{crit}} = 3\Mp^2H^2$ and $\Omega_{\mathrm{rad}, 0}h^2 = h^2\rho_{\mathrm{rad},0}/\rho_{\mathrm{crit}} \simeq 4.3\times 10^{-5}$.
In particular, we employ the publicly available code Cosmolattice~\cite{Figueroa:2020rrl,Figueroa:2021yhd} in a three-dimensional lattice, in order to numerically solve the metric perturbation equations in Eq.~\eqref{eq:Eqhij} along with the equations for the scalar system in an expanding universe.

\begin{figure}
\includegraphics[scale=0.55]{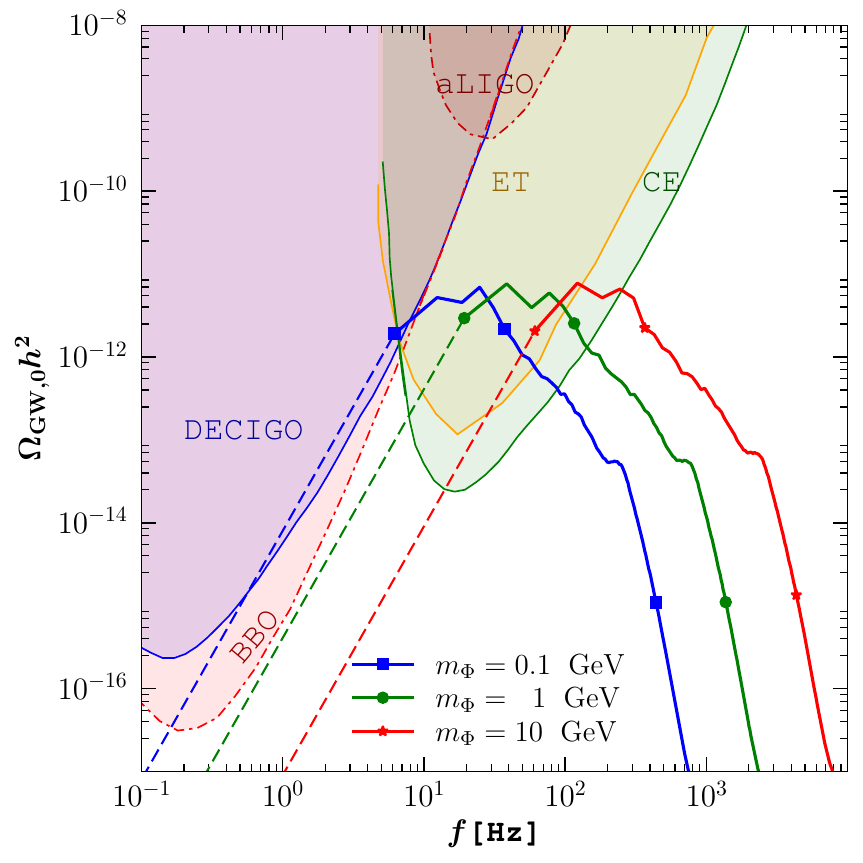}
\caption{We show the GW spectrum originating from the AD model considered here for three benchmark masses of $\Phi$. The potential parameters 
{$\lambda_\Phi\sim 10^{10}m_\Phi^2/M_{\rm {Pl}}^2$}, $A=0$ are chosen to ensure efficient parametric resonance (see the main text for details).
The fraction of initial scalar field energy density over the total energy density of the Universe is $\alpha=30\%$. The dashed lines at lower frequencies are extrapolated according to the causal superhorizon
 $k^3$ scaling.
}\label{fig:SpectraGWs}
\end{figure}

The spectra for three benchmark examples are shown in Fig~\ref{fig:SpectraGWs}.
For a $\Phi^4$-dominated system, as in our case initially, the field starts rolling/oscillating when $H\sim \sqrt{\lphi}\phiin$.
The combination $\sqrt{\lphi}\phiin$ also determines the frequency of the observed GWs.
We consider the RD Universe with the scalar field system contributing a fraction of the total energy $\alpha$; Fig~\ref{fig:SpectraGWs} corresponds to $\alpha = 0.3$. 
For GWs generated by sources peaked around a characteristic scale $k_c$ in the momentum space~\cite{Giblin:2014gra}, $\Omega^{\rm{ peak}}_{\rm{GW}}\propto\alpha^2\propto\phi^8_{\rm {in}}$, and we found that a sizable $\alpha\sim O(0.1)$ which optimizes GW amplitude requires $\phiin\sim \Mp$ {(see also~\cite{Cui:2023fbg})}.
Furthermore, the observed frequency today is $f_{\rm {GW}}^{\rm {peak}} \simeq 2.7\times 10^{10}  k_c / \sqrt{ H_c M_{\rm {Pl}}}$ Hz, {where $k_c\sim H\sim \sqrt{\lambda_{\Phi}}\phi_{\rm{in}}$}  (see Section E of Supplemental Material {for details}), {leading to $f_{\rm {GW}}^{\rm {peak}}\sim 10^{10}\lambda_\Phi^{1/4} $ for $\phi_{\rm {in}}\sim M_{\rm {Pl}}$.}
As we have seen from the discussion on the resonance structure of the model, the parameters $\mu_{\pm}$ (or simply $\mu$ when $A=0$) determine the nature and efficiency of resonance~\cite{Greene:1997fu}.
{We found that stable resonance occurs for $\mu\equiv {\mphi}/({\sqrt{\lphi}\phiin})\lesssim 10^{-5}$. We choose   $\mu = 10^{-5}$, which fixes the product $\lambda_\Phi \phi_{\rm{in}}^2$. With an optimal initial value of $\phi_{\rm{in}} \sim M_{\rm{Pl}}$, we find that $\lambda_\Phi\simeq 10^{-35}, 10^{-33}, 10^{-31}$ for $m_\Phi=0.1,1,10$ GeV; see Fig.~\ref{fig:SpectraGWs}.}
As discussed earlier, $m_\Phi\gtrsim O(0.1)$ GeV is required for $\Phi$ to successfully transfer the asymmetry to SM baryons. Interestingly, in the viable range of $m_\Phi\sim O(0.1)-O(10)$ GeV, the resulting peak GW frequency lies well within the range of ET/CE and with sufficient amplitudes, making signal detection promising for the near future.
A higher range of $m_\Phi$ is compatible with baryogenesis but requires futuristic, higher-frequency GW detectors \cite{Aggarwal:2020olq} to capture the peak frequency range of the signal.
Due to the finite box size and lattice spacing, low-frequency modes well off the peak cannot be resolved in our simulation, which is a common issue in the lattice simulation literature~\cite{Easther:2006vd}.
Therefore, to extend the spectral prediction to lower frequencies, we apply the standard assumption that at low wave numbers, the GW spectrum scales as $k^3$, as derived from causality arguments~\cite{Easther:2006vd, Caprini:2009fx, Cai:2019cdl, Hook:2020phx}. This is shown in dashed lines in Fig.~\ref{fig:SpectraGWs}. Depending on $m_\Phi$, the low-$k$ tail may be detectable by DECIGO and BBO, providing the prospect for correlated detection across multiple frequency bands, which simultaneously probes the peak and tail of the spectrum.
Before concluding, let us note that the simulated signal shown in Fig.~\ref{fig:SpectraGWs} is ${\cal O}(100)$ times larger than the lowest point of the CE sensitivity curve. Since $\Omega_{GW}h^2\propto \alpha^2$~\cite{Giblin:2014gra}, we can reduce $\alpha$, the initial fraction of the energy density of the scalar field with respect to the total energy density in the Universe, down to $\alpha\sim O(1\%)$ and still yield a GW signal within reach of CE.

\noindent\textbf{Conclusion and Outlook.}
In this Letter, we first demonstrate that the Affleck-Dine mechanism for baryogenesis can be realized by a complex light scalar field $\Phi$ with a mass $m_\Phi$ well below the electroweak scale without invoking supersymmetry.
In particular, we use a convex, renormalizable, polynomial potential (which does not lead to the formation of Q-balls, generally present in SUSY AD models).
Such a scalar field acts as a spectator field during inflation while acquiring a large field value at the end of inflation.
 The subsequent rolling down and oscillation of the scalar field occur long after inflation, during the radiation-dominated era, triggering the generation of baryon asymmetry. Parametric resonance and fragmentation generally arise during the oscillation of the scalar field, naturally leading to GW production. We demonstrate that the models considered can lead to successful baryogenesis while sourcing detectable GW signals with a peak frequency of O(10-100) Hz, within reach of experiments such as the ET and CE. Furthermore, the characteristic new physics scale, characterized by $m_\Phi$, intriguingly lies in the range of ${\cal O}(0.1\rm{-}10)$ GeV, while the transfer of the $\Phi$ asymmetry to the SM B- or L-asymmetry requires interactions between the new physics sector and the SM states. Hence, from this well-motivated scenario, a new, natural complementarity arises between SGWB detection and laboratory searches for new particle physics across the energy, intensity, and neutrino frontiers. The specifics of the complementary laboratory signal depend on the details of the asymmetry transfer mechanism. Examples include searches for ${\cal O}(0.1\rm{ -}10)$ GeV mass sterile neutrino with DUNE, SHiP, FASER, etc., and exotic multijet signals at the LHC (see Supplemental Material, Section D). This work thus sheds new light on probing early Universe baryogenesis through its GW imprint, potentially complemented by associated laboratory probes, while presenting a well-motivated new physics source for SGWB.

\medskip

\noindent\textbf{Acknowledgements.}
We thank Brian Shuve for helpful discussions.
YC is supported by the US Department of Energy under award number DE-SC0008541. AG is supported by "Excellence Initiative - Research University (2020-2026)" IDUB grant.
PS is supported by Grant-in-Aid for Scientific Research (B) under Contract No.~23K25873 (23H01177).
Numerical computation in this work was carried out at the Yukawa Institute Computer Facility.
EIS acknowledges support by the U.S.
Department of Energy, Office of Science, Office of High
Energy Physics program under Award Number DE-SC0022021.

\providecommand{\noopsort}[1]{}\providecommand{\singleletter}[1]{#1}%
%


\clearpage
\newpage

\onecolumngrid
\setcounter{secnumdepth}{3}
\setcounter{equation}{0}
\setcounter{figure}{0}
\setcounter{table}{0}
\setcounter{page}{1}
\makeatletter
\renewcommand{\theequation}{S\arabic{equation}}
\renewcommand{\thefigure}{S\arabic{figure}}
\renewcommand{\bibnumfmt}[1]{[#1]}
\renewcommand{\citenumfont}[1]{#1}
\pagestyle{plain}

\begin{center}
\Large{\textbf{Probing baryogenesis with gravitational waves}}\\
\medskip
\textit{Supplemental Material}\\
\medskip
{Yanou Cui,~Anish Ghoshal,~Pankaj Saha and Evangelos I.~Sfakianakis }
\end{center}

\subsection{Inflationary fluctuations and initial conditions}
\label{supp:ICs}

In some versions of the Affleck-Dine mechanism, an extra coupling is introduced in the potential in order to generate a radial minimum at large field values~\cite{Allahverdi:2012ju}. We argue that this is not necessary for the AD field to acquire a large initial value. It has been known that a light scalar field in de-Sitter space-time acquires quantum fluctuations, eventually leading to an equilibrium distribution~\cite{Starobinsky:1994bd}. In the case of a quartic real scalar field with $V= {1\over 4}\lambda\phi^4$, the variance of the equilibrium distribution is $\langle \phi^2\rangle \sim H^2/\sqrt{\lambda} $, whereas for a quadratic field $V={1\over 2}m^2\phi^2$ we arrive at $\langle \phi^2\rangle \sim H^4/m^2 $.
As an example,  $\langle \phi^2\rangle \sim M_{\rm {Pl}}^2$ and $0.1\, \rm{GeV} \lesssim m\lesssim 10 \, \rm{GeV} $  requires an inflationary Hubble scale of at least $10^{-10}M_{\rm{Pl}} \lesssim H \lesssim 10^{-9}M_{\rm{Pl}}$ for quadratic fields. If the quartic term dominates during inflation with $\lambda\sim 10^{-30}$ a Planckian standard deviation requires $H\sim 10^{-15} M_{\rm {Pl}}$ or greater. The variance of the distribution grows as $\langle \phi^2\rangle = (H^2/4\pi^2) N$ where $N$ is the $e$-folding number. We thus see that an equilibrium distribution with a Planckian standard deviation requires much more than the usual minimum amount of $60$ $e$-folds of inflation. This may be considered as a way to infer whether a prolonged inflationary phase took place in the early Universe. 

The above discussion (see Ref.~\cite{Cui:2023fbg} for more details) applies to the absolute value of the complex scalar field $|\phi|$. The argument will enter its equilibrium distribution much sooner, and the distribution itself will be much simpler: uniform between $0$ and $2\pi$. The intricacies of multi-dimensional random walks and their applications to de Sitter fluctuations of fields with continuous symmetries (including complex scalars) can be found in Ref.~\cite{Adshead:2020ijf}. 
Finally, a large initial field value $\phi_{\rm{in}}$ could simply result from an ad hoc initial condition. One may argue that the initial condition for the inflaton itself is set ad hoc, although quantum fluctuations can be invoked to explain it. 
Most importantly, our results are independent of the details of how $\phi_{\rm{in}}$ is realized (the same is true for the inflationary attractor). 

\subsection{Isocurvature perturbations}
\label{supp:isoc}
In our previous work~\cite{Cui:2023fbg}, we briefly discussed the possible isocurvature fluctuations from spectator scalar fields during inflation. In the current case, there are two degrees of freedom due to the complex nature of the AD field. If we decompose the complex scalar into a radial and an angular degree of freedom, the radial field will have similar behavior to a real scalar (for slight quantitative differences, see~\cite{Adshead:2020ijf}), with isocurvature fluctuations that scale as ${\cal P}_{\cal S} \sim H_I^2/M_{\rm {Pl}}^2$ if we take the initial value of the radial field to be ${\cal O}(M_{\rm {Pl}})$ originating solely form de-Sitter fluctuations as described in Section~\ref{supp:ICs}. We thus see that for reasonable values of the inflationary Hubble scale $H_I$, these fluctuations are suppressed.
Furthermore, if the AD field decays to SM particles, we can expect the isocurvature perturbation to vanish. 

However, the angular degree of freedom, which is responsible for the baryon number, also receives de Sitter fluctuations during inflation. This means that the AD mechanism proceeds differently in different parts of the Universe due to the fluctuations in the initial condition for the angular field. This leads to baryon isocurvature fluctuations, which are more constrained by the measurement of the CMB. 

We follow previous computations of isocurvature fluctuations in models of AD baryogenesis, in particular, Ref.~\cite{Hertzberg:2013mba}; see also~\cite{Kasuya:2008xp}.
During inflation, we can decompose the AD field $\Phi$ into a radial component $R$ and an angular component $\theta$. We consider a typical value of the radial field $\bar R = \sqrt{\langle R^2\rangle} \sim M_{\rm Pl}$ and thus
\beq
\langle \delta \theta^2\rangle \simeq {1\over 4\pi^2} {H_I^2\over \bar R^2}
\eeq
Ref.~\cite{Hertzberg:2013mba} introduced a multiplicative ${\cal O}(1)$ factor to account for the two-dimensional random walk of a complex scalar (see also Ref.~\cite{Adshead:2020ijf}). Since the baryon number is proportional to $\sin(n\theta)$ the normalized first order fluctuations $\delta n_b$ follow 
\beq
{\delta n_b\over n_b} = n \delta\theta \cot (\theta)
\eeq
The corresponding power of the baryon isocurvature temperature fluctuations is
\beq
\left \langle \left (
\delta T\over T
\right ) \right \rangle_{\rm {isoc}}
\simeq{0.004} {\Omega_b^2\over \Omega_m^2} {n^2 H^2\over \bar R^2} \cot^2(n\theta)
\eeq
We can take $\cot^2(n\theta)\sim 1$ and $\Omega_b/\Omega_m\simeq 0.16$, leading to 
\beq
\left \langle \left (
\delta T\over T
\right ) \right \rangle_{\rm {isoc}}
\simeq 10^{-4}{n^2 H^2\over \bar R^2} 
\eeq

Taking a simple model of single-field inflation for simplicity and following again Ref.~\cite{Hertzberg:2013mba}
\beq
\left \langle \left (
\delta T\over T
\right ) \right \rangle_{\rm {adiabatic}}
\simeq {1\over 20} {H^2\over 8\pi^2 M_{\rm {Pl}^2 }\epsilon}
={1\over 20} {H^2\over 8\pi^2 M_{\rm {Pl}^2 }\epsilon} \simeq 4 \times 10^{-5}
{H^2\over M_{\rm Pl}^2}{1\over r} 
\eeq
where we used the tensor-to-scalar ratio rather than the first slow roll parameter, using $r=16\epsilon$.
The ratio of the power in this isocurvature  over adiabatic perturbations is then
\beq
\alpha_{II} \sim n^2 r
\eeq
where we dropped ${\cal O}(1)$ terms and used $\bar R\sim M_{\rm {Pl}}^2$. We see that the Planck bound $\alpha_{II} < 3.9\times 10^{-2}$ is satisfied for $n=2$ and $r< 10^{-2}$, which is, for example satisfied for Starobinsky inflation, Higgs(-like) inflation and some $\alpha$-attractor models, where $r={\cal O}(10^{-3})$. As the inflationary scale is lowered, this bound is more easily satisfied.

\subsection{Estimate of the asymmetry}
\label{sec:asymmetryestimate}

We provide a more detailed derivation for the asymmetry calculation than the one found in the main text, following Refs.~\cite{Lloyd-Stubbs:2020sed,Lloyd-Stubbs:2022wmh}.
The scalar potential for the complex field  $V(\Phi) = \lf\absPhi^4 + \mf^2\absPhi^2 - A(\Phi^2 + \Phi_{\ast}^2)$ with quadratic symmetry-breaking terms is
\begin{equation}
    V(\Phi) = \lf\absPhi^4 + \mf^2\absPhi^2 - A(\Phi^2 + \Phi_{\ast}^2)
\end{equation}
The field equations for the real and imaginary parts are:
\begin{align}
\label{eqapp:phiR}
\ddot{\phi}_{_\mathrm{R}} &+ 3H\dot{\phi}_{_\mathrm{R}} + \mR^2\phiR + \lf(\phiR^2 + \phiI^2)\phiR = 0,\\
\label{eqapp:phiI}
\ddot{\phi}_{_\mathrm{I}} &+ 3H\dot{\phi}_{_\mathrm{I}}~ + \mI^2\phiI~ + \lf(\phiR^2 + \phiI^2)\phiI~ = 0,
\end{align}
where
\begin{align}
\mR^2 = \mf^2 - 2A;\quad \mI^2 = \mf^2 + 2A.
\end{align}
while the Hubble expansion is that of a radiation-dominated Universe ($H\propto t^{1/2}$.)
\par
At large field values (defined as $|\Phi|>\phi^*=m_\Phi / \sqrt{\lambda_\Phi}$), the symmetry-breaking term is subleading, and we consider the dominant term of the potential 
\begin{equation}
    V(\Phi) \simeq \lambda_\Phi |\Phi|^4\,,~~~ |\Phi|>\phi_*,\end{equation}
    while at small field values, the quartic term is negligible, and we may approximate the potential as
\begin{equation}V(\Phi) \simeq m_\Phi^2 |\Phi|^2-A\left (\Phi^2 + (\Phi^\ast)^2\right ) \,,~~~ |\Phi|<\phi_*
\end{equation}

 By defining time $t_*$ as
$|\Phi(t_*)| =\phi_*$, we approximate the field evolution during the  two eras as~\cite{Turner:1983he}:
 \par\noindent
 When $t<t_\ast$:
 \begin{eqnarray}
     \phi_{\rm R}(t<t_\ast) &=& \left(\frac{a_{\rm in}}{a}\right)\phi_{\rm R,in}
     \label{eq:phiReBast}
     \\
     \phi_{\rm I}(t<t_\ast) &=& \left(\frac{a_{\rm in}}{a}\right)\phi_{\rm I,in}
     \label{eq:phiImBast}
     \\
     \phiR(t>t_\ast) &=& \phiRS\left(\frac{a_{\ast}}{a}\right)^{3/2}\cos(\mR(t-t_{\ast})),
     \label{eq:phiReast}
     \\
     \phiI(t>t_\ast) &=& \phiIS~\left(\frac{a_{\ast}}{a}\right)^{3/2}\cos(\mI(t-t_{\ast}))
     \label{eq:phiImast}
 \end{eqnarray}
where $\phiRS = (a_{\rm {in}}/a_* )\phi_{\rm R,in}$ and 
 $\phiIS = (a_{\rm {in}}/a_* )\phi_{\rm I,in}$.
The conserved charge density in the $\Phi$ condensate is
\begin{align}
    n_0(t) = i\left(\Phi^{\dagger}\dot{\Phi} - \dot{\Phi}^{\dagger}\Phi\right)
         = \dot{\phi}_{\rm R}\phiI - \phiR\dot{\phi}_{\rm I}
\end{align}
where we have used the subscript `$0$' to denote the generated asymmetry in the $\Phi$ sector alone, without considering the decay to SM particles. 

Substituting the approximate solutions for the background fields, the asymmetry becomes
\begin{equation}
    n_0(t) = \phiRS\phiIS\left(\frac{a_{\ast}}{a}\right)^{3}\left[\mI\sin(\mI(t-t_\ast))\cos(\mR(t-t_\ast)) - \mR\sin(\mR(t-t_\ast))\cos(\mI(t-t_\ast))\right] \, .
\end{equation}
Since the symmetry-breaking term is subdominant to the other terms in the potential, we consider  $A\ll \mf^2$ and expand the asymmetry $n_0(t)$ at leading order in $A/\mf^2$
\begin{equation}
    n_0(t) = m_\Phi \phiRS\phiIS\left(\frac{a_{\ast}}{a}\right)^{3}\left[\sin\left(\frac{2A(t-t_\ast)}{\mf}\right) + \frac{A}{\mf^2}\sin(2\mf(t-t_\ast))\right] \, .
    \label{eq:asym0}
\end{equation}
The second term is highly oscillatory and will amount to zero when averaged over several field oscillations. Hence, for $t>t_\ast$ the expression in Eq.~(\ref{eq:asym0}) reduces to 
\begin{equation}
    n_0(t) = \phiRS\phiIS\left(\frac{\phiIn}{\phiS}\right)\left(\frac{a_{\rm in}}{a}\right)^{3}\mf\sin\left(\frac{2A(t-t_\ast)}{\mf}\right) \, .
    \label{eq:asym}
\end{equation}
The generated asymmetry will oscillate around zero with a period $T_{\mathrm{asy}}=\pi\mf/A$.
Defining the comoving asymmetry $\nc(t)\equiv (a(t)/a_{\rm in})^3n(t)$, we can estimate the transfer of generated asymmetry to a conserved SM asymmetry via $B$-conserving $\Phi$ decay to SM particles with a constant rate $\Gphi$.

Finally, the transferred asymmetry to the SM can be calculated as
\begin{equation}
    \nct(t) = \int_{t_\ast}^t {\rm d}t~\Gphi n_{_{\mathrm{C}}}(t)
         = \frac{\Gphi\phiS^2\sin(2\theta)\mf}{2}\int_{t_\ast}^t {\rm d}t~e^{-\Gphi(t-t_\ast)}\sin\left(\frac{2A(t-t_\ast)}{\mf}\right)
         \label{eq:ncintegral}
\end{equation}
with $\phiRS\phiIS = \phiS^2\sin\theta\cos\theta = \frac{1}{2}\phiS^2\sin 2\theta$. 
The final expression for the transferred asymmetry for $t-t_* \gg m_\Phi/A$ is 
\begin{align}
    \nct= \frac{\Gphi\phiS^2\sin(2\theta)\mf^2}{2A}\left(1 + \left(\frac{\Gphi\mf}{2A}\right)^2\right)^{-1}.
    \label{eq:asym_exp}
\end{align}
This expression has two limiting cases, depending on whether the $\Phi$ lifetime 
 is short ($\Gphi \gg 2A/\mf$) or long ($\Gphi\ll 2A/\mf$) compared to the period of oscillation of the asymmetry:
\begin{align}
    \nct = 
    \begin{cases}
    \frac{2A}{\Gphi} \, \phiS^2 \sin(2\theta) \quad 
    & \Gphi \gg 2A/\mf \\
    \frac{\Gphi  \mf^2}{2A}\,  \phiS^2 \sin(2\theta)  \quad 
 & \Gphi\ll 2A/\mf
    \end{cases}
    \, .
    \label{eq:asymmetrycasesappendix}
\end{align}

In the absence of efficient washout (as discussed in the main text and Supplemental Material Sect.~\ref{sec:transfer}), $\Phi$ asymmetry $\nct/s$ corresponds directly to baryon or lepton number asymmetry, which we assume below for simplicity. In case of a large washout effect, $\nct/s$ would be larger than the resultant $n_B/s$. We do not elaborate on how the washout affects the prediction in $n_B/s$ in the latter case, as it is highly dependent on the details of the asymmetry transfer mechanism. Instead, we will demonstrate the feasibility of generating $\nct/s\gtrsim10^{-10}$ that can compensate for a potentially large washout suppression. 
\\
Ultimately,  the quantity $n_B/s$ needs to match  the observed value $$\frac{n_B}{s}\Bigg|_{\mathrm{obs}} = 0.861\pm0.005\times 10^{-10}.$$
To this end, we multiply the comoving transferred asymmetry by $(a_{\ast}/a)^3$ to compute the total asymmetry
\begin{align}
    \tilde{n} = \left(\frac{a_\ast}{a}\right)^3\nct
\end{align}
For the RD Universe, as in our case, the equation of state is $w=1/3$, leading to 
\begin{align}
    \left(\frac{a_\ast}{a}\right)^3 =  \left(\frac{H(t)}{H_{\ast}}\right)^{3/2}
\end{align}
Dividing $\tilde n$  by the entropy density $s = 4k_T^2T^3$, with $k_T = \sqrt{\pi^2g(T)/90}$, we arrive at the baryon-to-photon ratio
\begin{align}
    \frac{n_B}{s} = 
    \begin{cases}
    \left(\frac{4\alpha^3}{k_{T_d}^6\lf}\right)^{1/4}\left(\frac{A}{\mf^2}\right)\left(\frac{\mf\Mp}{T_d^2}\right)\sin(2\theta)\quad &\Gphi \gg 2A/\mf\\
    \left(\frac{\alpha^3k_{T_d}^2}{64\lf}\right)^{1/4}\left(\frac{\mf^2}{A}\right)\left(\frac{T_d^2}{\mf\Mp}\right)\sin(2\theta)\quad &\Gphi \ll 2A/\mf \, ,
    \end{cases}
\end{align}
where   the $\Phi$ decay  completes when $\Gphi=H(T_{d})  = k_{T_d}T_d^2/\Mp$ and the Hubble scale at $t_*$ is $H_{\ast} \simeq \frac{3}{4}{\mf^2}/{\lf}$. 
We can also elucidate the dependence on $\phi_{\mathrm{in}}$, by rewriting the above expression as
\begin{align}
    \frac{n_B}{s} = 
    \begin{cases}
\mu\left(\frac{4\lf\alpha^3}{k_{T_d}^6}\right)^{1/4}\left(\frac{A}{\mf^2}\right)\left(\frac{\phi_{\mathrm{in}}\Mp}{T_d^2}\right)\sin(2\theta)\quad &\Gphi \gg 2A/\mf\\
    \mu\left(\frac{\alpha^3k_{T_d}^2}{64\lf^3}\right)^{1/4}\left(\frac{\mf^2}{A}\right)\left(\frac{T_d^2}{\phi_{\mathrm{in}}\Mp}\right)\sin(2\theta)\quad &\Gphi \ll 2A/\mf \, ,
    \end{cases}
    \label{eq:nb_s}
\end{align}
where for sufficient resonance, we require $\mu^2 \equiv \mf^2/(\lf\phi_{\mathrm{in}}^2) \lesssim 10^{-5}$.

When written in terms of the dimensionless parameters, $\epsilon_\Phi = A/m_\Phi^2$ and $\gphi = \Gamma_{\Phi}/\mphi$, we find the following familiar expression:
\begin{align}
    \frac{n_B}{s} =
    \begin{cases}
    \left(\frac{4\alpha^3}{\lphi k_{T_d}^6}\right)^{1/4} \epsilon_\Phi\frac{\mphi M_{\rm {Pl}}}{T_d^2}\sin(2\theta);&\gphi \gg 2\epsilon_\Phi\\
  \left(\frac{\alpha^3k_{T_d}^2}{64\lphi}\right)^{1/4}\frac{1}{\epsilon_\Phi}\frac{T_d^2}{\mphi M_{\rm{Pl}}}\sin(2\theta); &\gphi \ll 2\epsilon_\Phi
    \end{cases}
    \label{eq:nbsmainDmnlessApp}
\end{align}

\begin{figure}[h!]
\centering
\includegraphics[width=0.95\textwidth]{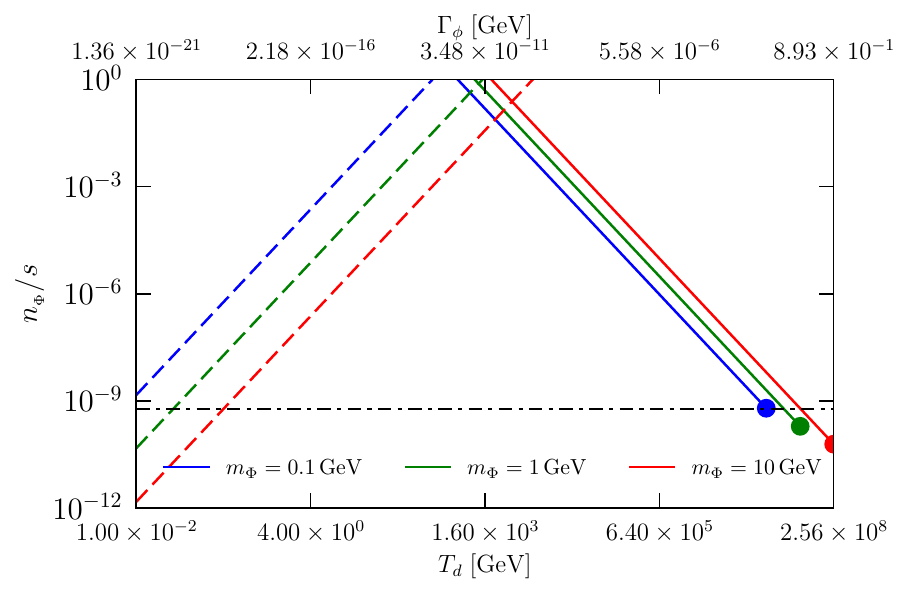}
\caption{The maximum baryon-to-photon ratio
{$\left . {n_B / s}\right|_{\rm {max}} = {n_\Phi/s}$} (neglecting possible washout and the sphaleron factor) for three different masses of the scalar field, as in Fig.~(1) of the main text, as a function of the temperature $T_d$ of the $\Phi$ decay. 
The solid and dashed lines correspond to $\epsilon_{\Phi} = 10^{-20}$ and $\epsilon_{\Phi} = 10^{-6}$ respectively.
The maximum temperature (denoted as the dots) for each line corresponds to the case of instantaneous decay of the scalar at $t=t_{\ast}$. 
The horizontal dot-dashed line corresponds to the observed value of the baryon asymmetry.
}
\label{fig:asymmetry1}
\end{figure}

Figure~\ref{fig:asymmetry1} shows the resulting asymmetry, following Eq.~\eqref{eq:asym_exp}, as a function of the temperature of the Universe at the time of complete $\Phi$ decay for different values of the potential parameters; $m_\Phi$ and $A$. The quartic term is chosen to keep $m_\Phi^2/\lambda_\Phi\phi_{\mathrm{in}}^2 = 10^{-5}$ for sufficient resonance.  We see the strong dependence of $n_B/s$ on $A$ and $T_d$, as expected. This enables the model to remain viable, even in cases where significant washout occurs, by increasing the initially produced asymmetry through appropriate parameter choices.
Finally, let us reiterate one assumption made in the above analytic results regarding the generated asymmetry. As mentioned in the main text, our analysis assumes that the motion of $\Phi$ is radial at early times, for $\lambda_\Phi |\Phi|^4 \gg m_\Phi^2 |\Phi|^2$. After that, the potential is dominated by the quadratic terms, which in our case include the bare mass and the asymmetry $A(\Phi^2 +\Phi_*^2)$. By simply equating the quartic and bare mass terms, we arrive at the time of transition between the two regimes of $t_*$ when $|\Phi(t_*) |= m_\Phi / \sqrt{\lambda_\Phi}$. Demanding that at early times (before $t_*$) the asymmetry term can be neglected leads to $\epsilon_\Phi < (\mu^2/15\alpha)(\phi_{\rm {in}}^2/M_{\rm {Pl}}^2)$ where $\mu^2\equiv m_\Phi^2 / (\lambda_\Phi \phi_{\rm {in}}^2) \lesssim 10^{-5}$. For $\phi_{\rm {in}} \sim M_{\rm {Pl}}$ the above inequality can be written as 
$\epsilon_\Phi \lesssim 10^{-5}$. If we push $\epsilon_\Phi$ above this value, the model is not invalidated, but the asymmetry would get an extra suppression factor since $\sin\theta < \sin\theta_{\rm {in}}$. The amount of suppression depends on the amount of non-radial motion for $t<t_*$. That being said, the computed $n_B/s$ can be many orders of magnitude above the measured value of $10^{-10}$, meaning that a suppression of $\sin\theta$ is not ``fatal'' for the model, even for $\epsilon_\Phi>10^{-5}$. We do not perform a detailed analysis for this regime, which involves simply computing the angular motion of $\Phi$ from the first instant when it starts rolling, given a specific set of potential parameters. 

As an example, for the benchmark values of $\lphi\sim 10^{-30}$ and $\mphi=10\,\mathrm{GeV}$, we must have $\epsilon_\Phi \lesssim 10^{-6}$ for the validity of our estimates in Eq.~(\ref{eq:nb_s}). 
The temperature at the decay can vary between $T_d^{\mathrm{max}}\sim 10^8\,\mathrm{GeV}$ for instantaneous decay and $T_d^{\mathrm{min}}=10\mathrm{MeV}$, if the decay occurs right before BBN~\cite{Kawasaki:1999na,Kawasaki:2000en,Steigman:2007xt,Hasegawa:2019jsa}.
Therefore, to obtain the correct ratio of $n_B/s$, we require $\epsilon_\Phi = 10^{-38}(T_d/\mathrm{GeV})^2$ when $\gphi \ll \epsilon_\Phi$. 

\subsection{Transfer of the $\Phi$ asymmetry to the SM baryon asymmetry}
\label{sec:transfer}
In this section, we elaborate on the example models realizing the transfer of $\Phi$ asymmetry to the SM baryon asymmetry and the related constraints.\\

\noindent \textit{(A) $\Phi$ decay to leptons (leptogenesis).}\\
A simple renormalizable model involves:
\beq
\mathcal{L}\supset y_{N,~ i}\Phi\bar{N_i}^C{N_i}+g_{\nu,~ij}\bar{N}_{i}HL_j+ {\rm h.c.} \label{Eq: MajoranaN}
\eeq
where $N_i$'s are right-handed (RH) Majorana right-handed neutrinos, $H$ is SM Higgs, $L_j$ are SM lepton doublets. The number of $N$ species, $N_N$, is a model-dependent parameter, and one can take $N=1$ for simplicity or $N_N=3$ as an example by analogy to the SM neutrinos. Again, while it is appealing to simultaneously address the origin of neutrino masses, it is not imposed as a requirement for our consideration. 

We consider the simple scenario where $\Phi$ decays to $NN$, then $N$ transfers the asymmetry to SM leptons via freeze-out or freeze-in processes as considered in leptogenesis literature \cite{Akhmedov:1998qx, Klaric:2020phc, Klaric:2021cpi, Cui:2011qe, Flood:2021qhq, Shuve:2014zua, Hambye:2016sby}. As seen from our GW results, $m_\Phi\lesssim O(10)$ GeV is preferred in order to yield a detectable GW frequency below $\sim100$ Hz. This implies that light $N$ of mass $\lesssim O(1-10)$ GeV. In this mass range, conventional $N$ decays to SM leptons via a 3-body or loop-suppressed process, which typically occurs after electroweak sphaleron already turns off, albeit before BBN. The washout effect from inverse decay or backscattering is also significant in this low-mass range. However, as found in recent leptogenesis literature, asymmetry transfer from $N$ to SM leptons can be realized while sphaleron is still active by freeze-out (generalized out-of-equilibrium decay or scattering) or freeze-in (via $N$-$\nu_{\rm SM}$ mixing/oscillation) mechanisms, for $m_N$ as low as O(100) MeV. The strength of the washout effect can be highly flavor-dependent. Even in the presence of large washout suppression, as shown in the main text, our Affleck-Dine model can initially generate a larger $\Phi$ asymmetry, which compensates for the washout to eventually result in the observed baryon asymmetry. Unlike with the leptogenesis case, here, the $NHL$ interaction only serves the purpose of transferring the $\Phi$ ($N$) asymmetry to SM leptons instead of generating the asymmetry via CP violation in the SM neutrino sector. Therefore, the details of the asymmetry evolution in our case may differ from those of its leptogenesis counterparts and are highly model-dependent, which can be elucidated by dedicated studies beyond the scope of this work. Furthermore, an alternative to the Majorana $N$ as shown in Eq.~\ref{Eq: MajoranaN} is to introduce quasi-Dirac $N$, where the washout effect can be more suppressed for the GeV mass range (also see \cite{Cline:2020mdt} for this possibility and the related phenomenological implications).
 For all of the above possibilities for $N$ to SM asymmetry transfer, $\Phi\rightarrow NN$ decay is required to occur before EW sphaleron turns off, i.e. $\Gamma_{\Phi\rightarrow NN}\sim y_\Phi^2m_\Phi/(16\pi)\gtrsim H(T_{\rm EW})$. This leads to the constraint $y_\Phi\gtrsim10^{-7}\sqrt{\frac{\rm {GeV}}{m_\Phi}}$, which can be easily satisfied.

 Laboratory signatures of the case of $\Phi$ decaying to leptons generally relate to sterile neutrino searches in the mass range of $O(0.1-10)$ GeV, which are being conducted at a range of experiments such as DUNE, SHiP, FASER et al. \cite{sterilenubounds, Bolton:2019pcu,FASER:2018eoc,Ballett:2019bgd,T2K:2019jwa,Drewes:2018gkc,Barouki:2022bkt}.

\noindent \textit{(B) $\Phi$ decay to baryons.}\\
A simple renormalizable model in this case includes:
\beq
\mathcal{L}\supset y_\chi\phi_\chi\chi\chi+g_i\chi\bar{u}_i\varphi+\lambda_{ij}\varphi d_i d_j +{\rm h.c.}
\eeq
where $\chi$ is a singlet Majorana fermion, $\varphi$ is an up-type diquark scalar, and the flavor indices $i,~j=1,2,3$ in down-type quark combination $d_i d_j$ should be antisymmetric.
This type of model finds its close analogy in supersymmetric theories. In particular, $\chi$ may relate to a neutralino or singlino \cite{Cui:2012jh}, $\varphi$ to an up-type squark, and $\lambda_{ij}$ to a UDD-type R-parity-violating (RPV) coupling. Therefore, existing searches for RPV SUSY may lead to constraints on masses and couplings that we consider in this context. The LHC constraints on the colored particle $\phi$ depend on whether it resembles the first two generations of squark or stop, and depend on whether RPV decay $\phi\rightarrow dd$ or R-parity conserving decay $\phi\rightarrow u\chi$ dominates (assuming $\chi$ reveals itself as MET) \cite{Karmakar:2023mhr, CMS:2018mts, CMS:2021beq, ATLAS:2020syg}. In any case, generally, $m_\varphi\gtrsim $1 TeV is the ballpark of the current LHC constraints, which we will use for our estimates. This constraint determines that in the parameter range of our interest, where $m_\chi\sim O(1-10)$ GeV, $\chi$ decays proceed with the 3-body channel $\chi\rightarrow udd$. For simplicity, we assume $g_i\sim1$. B-violating coupling $\lambda_{ij}$ is subject to various constraints by laboratory searches at the intensity frontier. This model does not involve L-violation and does not induce proton decay, which usually introduces constraints on B-violating couplings \cite{Arnold:2012sd}. Due to the antisymmetry in $i,~j$, $n$-$\bar{n}$ constraint is rather weak, while di-nucleon decay $pp\rightarrow K^+K^+$ provides the strongest bound \cite{Goity:1994dq, Barbier:2004ez} of $\lambda_{12}\lesssim10^{-6}$, for $m_\varphi\sim$ TeV, $m_\chi\sim$ O(GeV). However, the bound is relaxed for heavy flavor quarks, e.g., $\lambda_{23}\lesssim 1$. 

The kinematic condition of $\chi$ decaying to baryons leads to a lower limit on masses: $m_\Phi> 2$ GeV, $m_\chi>1$ GeV. The decays of both $\Phi\rightarrow \chi\chi$ and $\chi\rightarrow udd$ are required to occur before BBN, i.e., $\Gamma_{\Phi\rightarrow \chi\chi}\sim\frac{1}{16\pi}y_\chi^2m_\Phi> H(T_{\rm {BBN}})$, and $\Gamma_{\chi\rightarrow udd}>H(T_{\rm {BBN}})$. We then estimate the constraint on coupling parameters as: $y_\chi\gtrsim10^{-12}\frac{\rm GeV}{m_\phi}$, 
$\sum_{i, j} g_i\lambda_{ij}\gtrsim\sqrt{10^{-21}\frac{m^4_\varphi\rm {GeV}}{m_\chi^5}}$. 
With $m_\varphi\sim$ TeV, $m_\chi\sim$ O(GeV), we find that $\sum_{i, j}g_i\lambda_{ij}\gtrsim10^{-5}$. If $\lambda_{12}$ dominates $\chi$ decay, this would contradict the aforementioned di-nucleon decay bound. 
Therefore, we choose to consider the pattern of heavy flavor domination, which is well-motivated and has been widely studied in the SUSY context \cite{Brust:2011tb, Brust:2012uf, Cui:2011qe, Franceschini:2012za}. In particular, we consider the dominant decay of $\chi$ involves b-quark $\chi\rightarrow udb$ or $\chi\rightarrow usb$. As $m_b\simeq4$ GeV, the kinematics requires $m_\chi>$4 GeV and $m_\Phi>$8 GeV. The involvement of heavy b-quark also alleviates the potential washout suppression of baryon asymmetry due to inverse decay $udd\rightarrow \chi$ and back-scattering $udd\rightarrow\bar{u}\bar{d}\bar{d}$: by requiring $\chi$ decay below $T\sim m_b$ (compatible with the condition of decay before BBN), the washout effect gets Boltzmann suppression and quickly becomes inefficient. 

Laboratory signatures of the case of $\Phi$ decaying to baryons depend on model details. One interesting possibility is a multi-jet signal at the LHC (eight jets with b-tagging), potentially with displaced vertices, from pair-production of $\phi$ followed by $\phi\rightarrow\chi u\rightarrow uudd$, which is distinct yet relates to existing searches such as \cite{ATLAS:2018umm}.

\subsection{Parametric resonance and asymmetry dependence of GW production}
\label{sup:resonance}

It is worth exploring in more detail the connection between the asymmetry of the potential and the shape and amplitude of the resulting GW spectrum. 
The growth of fluctuations and their backreaction breaks the homogeneous oscillation of $\Phi$, leading to scalar field inhomogeneities with an effective anisotropic stress tensor $\Pi_{ij}^{TT} = \nabla_i\phi^b\nabla_j\phi^b$ (where the repeated index $``b"$ indicates the number of the scalar components and a summation is implied).
These time-dependent field inhomogeneities source (sub-Hubble) gravitational waves whose frequency depends on the characteristic energy scale involved $H_c$. 
The peak amplitude and frequency of GW spectra generated from sources peaked at some characteristic scale $k_c$ in momentum space can be estimated as~\cite{Giblin:2014gra}
\begin{align}
\label{eq:gt_freq}
f_{\rm {GW}}^{\rm {peak}} =& 2.7\times 10^{10} \sqrt{\frac{H_c}{M_{\rm{Pl}}}}\frac{k_c}{H_c}\, {\mathrm{Hz}} \, ,
\\
\Omega_{\rm {GW}}^{\rm {peak}} =& 2.3\times 10^{-4} \, \alpha^2\, \beta\, w 
\left ( \frac{k_c}{\sigma} \right ) 
\left (\frac{H_c}{k_c}\right )^2 \, ,
\label{eq:gt_Omega}
\end{align}
where $\alpha$ is a measure of the fraction of the energy in the GW source -- here the scalar field inhomogeneities -- relative to the total energy density of the Universe, and $\beta$ encodes the degree of anisotropy of the source, which must be determined from simulations. 
In typical scenarios $\beta=\mathcal{O}(0.01-0.1)$.
The equation of the state of the Universe at the time of the production of GWs for a radiation-dominated Universe is $w=1/3$ and 
$\sigma$ is the width of the source in momentum space, which can be taken as $\sigma\sim k_c$ for peaked sources (see Ref.~\cite{Giblin:2014gra} for the derivation and e.g. Refs.~\cite{Cui:2021are, Cui:2023fbg} for further applications of Eqs~\eqref{eq:gt_freq}, \eqref{eq:gt_Omega}).

Let us start from the symmetric case $A=0$, where the potential becomes 
\begin{equation}
    V(\phi_R,\phi_I) = \frac{1}{4} \lambda_\Phi(\phi_R^2 + \phi_I^2)^2 +  \frac{1}{2} m_\Phi^2 (\phi_R^2 + \phi_I^2)
    =  \frac{1}{4} \lambda_\Phi \phi_R^4 +  \frac{1}{4} \lambda_\Phi \phi_I^4 + 
    \frac{1}{2}   m_\Phi^2 \phi_R^2 + \frac{1}{2}  m_\Phi^2 \phi_I^2
     +  \frac{1}{2}  \lambda_\Phi \phi_R^2\phi_I^2
\end{equation}

Since the potential is radially symmetric, we can redefine the fields as $\phi_\parallel = \phi_R \cos\theta - \phi_I \sin\theta$ and $\phi_\perp = \phi_R \sin\theta + \phi_I \cos\theta$, where at the background level $\bar\phi_\perp=0$. This essentially follows our earlier work~\cite{Cui:2023fbg}, where we considered (among others) systems with two scalar fields featuring a quartic coupling, with one field initially displaced from its minimum and the other (classically) at zero. This specific example combines Models A and D from Ref.~\cite{Cui:2023fbg}. The equations of motion for the fluctuations in the $\phi_\parallel-\phi_\perp$ basis are
\begin{figure}[!ht]
    \centering
   \includegraphics[width=0.95\linewidth]{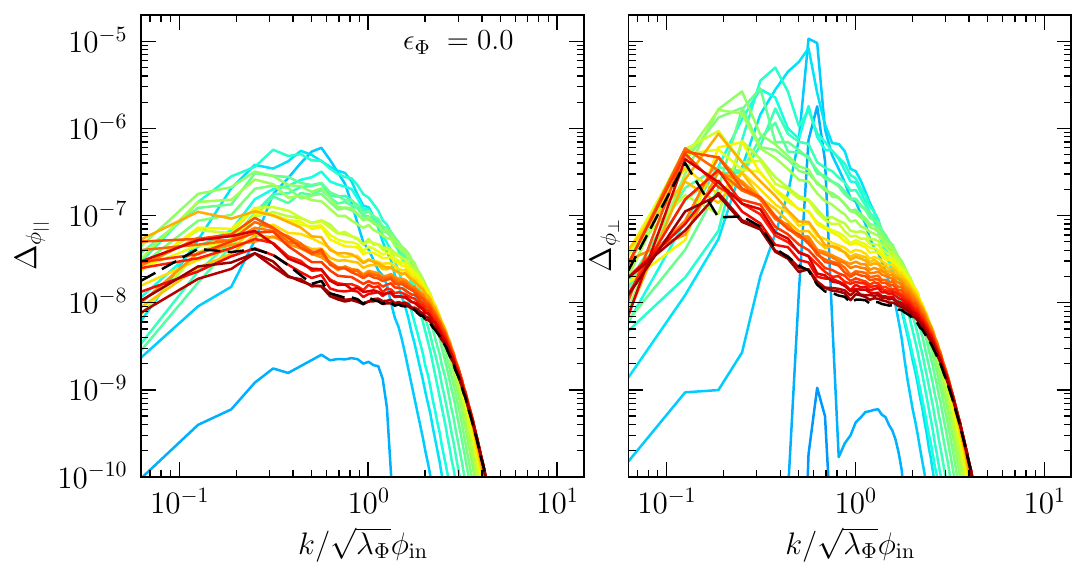}
    \caption{
   The spectra of the scalar field fluctuations along (left) and perpendicular (right) to the direction of the background trajectory in the case when the potential has a rotation symmetry with $A=0$, $\lambda_\Phi = 5\times 10^{-35}$ and $m_\Phi = 0.1$ GeV.
   }
    \label{fig:PSp_phichi}
\end{figure}
\begin{eqnarray}
\ddot{ \delta\phi_\parallel} + 3H \dot{\delta\phi_\parallel }+ \left ({k^2\over a^2}+ m_\Phi^2  + 3\lambda_\Phi \bar\phi^2 \right )\delta\phi_\parallel=0
\\
\ddot{ \delta\phi_\perp }+ 3H \dot{\delta\phi_\perp } + \left ({k^2\over a^2}+ m_\Phi^2  + \lambda_\Phi \bar\phi^2 \right ) \delta\phi_\perp  =0 \label{eqA: deltas}
\end{eqnarray}
In principle, we can see two sources of parametric resonance: self-resonance of the $\phi_\parallel $ field due to the quartic term $3\lambda_\Phi \bar\phi^2\delta\phi_\parallel $ and resonant amplification of the $\phi_\perp $ field through the coupling term $\lambda_\Phi \bar \phi^2 \delta\phi_\perp$. 

The background field trajectory is (by construction) along $\bar\phi_\perp =0$ and thus a straight line through the origin on the $\phi_R-\phi_I$ plane. Floquet analysis of a quartic field in a nonexpanding universe predicts that the resonance is stronger for $q=1$~ ($\mu_k^{max}\sim 0.15$) than $q=3$ ($\mu_k^{max}\sim 0.04$)~\cite{Greene:1997fu}, where $q$ is defined as the coefficient before $\lambda_\Phi \bar\phi^2$ in Eqs.~\ref{eqA: deltas}. Figure~\ref{fig:PSp_phichi} shows the spectra of $\delta\phi_\parallel,\delta\phi_\perp$ for different times, where we see that the $\delta\phi_\perp$ spectrum dominates, as expected 
from the standard analysis of preheating of quartic fields~\cite{Greene:1997fu}.
In the $\phi_R-\phi_I$ basis, the spectra will be identical for $\theta=\pi/4$, since each spectrum has an equal contribution from $\delta\phi_\parallel$ and $\delta\phi_\perp$.
 If the initial angle between the real and imaginary parts of $\Phi$ changes, the spectra of $\delta \phi_\parallel$ and $\delta\phi_\perp$ remain unchanged, whereas the spectra of $\phi_R$ and $\phi_I$ change, since the inner product between the basis vectors of the two bases changes with the angle $\theta$. That being said, the resulting GW spectrum is unchanged, as expected, due to the rotational symmetry of the potential.

\medskip

\begin{figure}[!ht]
\centering
\includegraphics[width=0.95\textwidth]{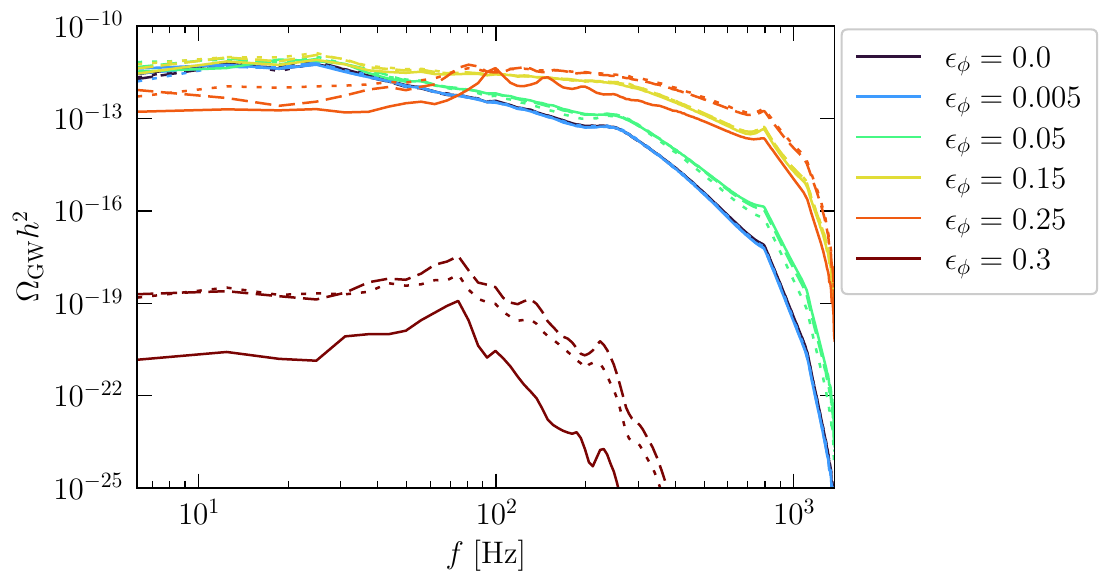}
\caption{ The resulting GW spectra for different values of the asymmetry parameter $A$ (color-coded according to the legend) and initial angle $\theta\equiv \arctan(\phi_I/\phi_R)|_{\rm{init}} = \pi/6, \pi/4, \pi/3$ (solid, dashed and dotted respectively).
}
\label{fig:compareA}
\end{figure}

We now move to the case of $A\ne 0$. We parametrize the effect by $\epsilon_\Phi=A/m_\Phi^2$.
Let us first discuss the background trajectory. For $\epsilon_\Phi=0$ it is a straight line segment through the origin, with a length of ${\cal O}(\phi_{\rm{in}})$ which will shrink in time due to red-shifting. For $0<\epsilon_\Phi\ll 1$, the trajectory during each period will resemble an ellipse. Using Eqs.~\eqref{eq:phiReast}-\eqref{eq:phiImast}) and Taylor-expanding for $\epsilon_\Phi\ll 1$, we can see that the minor axis of the (approximate) ellipse scales as ${\cal O}( \phi_{\rm {in}}\cdot A/m_\phi^2)$, whereas the major axis is simply ${\cal O} (\phi_{\rm {in}})$. The ellipticity is then $\sim (A/m_\phi^2)\ll 1$ for all cases of interest shown in the main text. For small values of $\epsilon_\Phi$, the background trajectory will be an ellipse with a very small ellipticity, and it will therefore behave almost like a straight line segment. We thus expect the parametric resonance results to be unaffected by the exact value of $\epsilon_\Phi\ll 1$.
As seen in Fig.~\ref{fig:compareA}, for $\epsilon_\phi \le 0.005$, the GW spectrum is independent of the initial angle $\theta_{\rm {in}}$ and indistinguishable from the symmetric case with $\epsilon_\Phi=0$. This means that for the asymmetry values that we use for our benchmark examples $\epsilon_\Phi <10^{-5}$, the resonance is ``blind'' with respect to the exact value of asymmetry in the potential. 
That being said, it is worth pushing the asymmetry value higher in order to cover the entire possible parameter space. For $0.05\lesssim \epsilon_\Phi \lesssim 0.15$, we see an enhancement of the GW spectrum at large values of the frequency $f\gtrsim 100$ Hz, while the independence of $\theta_{\rm {in}}$ still holds. This is an intriguing observation, as it provides an additional motivation for GW detectors that can probe frequencies around $1$ kHz with a sensitivity similar to CE and ET~\cite{Zhang:2022yab, Chen:2023axy}. For $\epsilon_\Phi\simeq 0.25$, the UV enhancement of the GW spectrum remains, but we also see a difference between the initial angle choices between the real and imaginary parts of $\Phi$. This means that a possible detection of a spectrum with such a UV enhancement would suffer from a partial degeneracy between $\epsilon_\phi$ and $\theta_{\rm {in}}$. 
Since we focused on $\epsilon_\Phi \ll 1$, such a highly model-dependent calculation is outside the scope of the present work. Finally, for $\epsilon_\Phi\gtrsim 0.3$, parametric resonance is severely suppressed, rendering the GW signals below the detection threshold.

Before concluding, we must note that since $\Phi$ is coupled to SM fermions in order to transfer the asymmetry to baryons, the effect of parametric excitation of these fermionic states should be evaluated. Nevertheless, in the context of preheating, it was found that during scalar field oscillations, the energy transfer to bosons is typically faster and larger than to fermions due to Pauli blocking \cite{Greene:1998nh, Giudice:1999fb, Greene:2000ew, Sorbo:2001xx, Adshead:2015kza}. Therefore, we expect our resonance analysis to remain robust in the presence of fermion couplings.


\end{document}